\documentclass[aps,pre,twocolumn,superscriptaddress,
floatfix,showkeys]{revtex4}
\usepackage{color,graphicx}
\newcommand{\beq}{\begin{equation}}
\newcommand{\eeq}{\end{equation}}
\newcommand{\bea}{\begin{eqnarray}}
\newcommand{\eea}{\end{eqnarray}}
\newcommand{\bean}{\begin{eqnarray*}}
\newcommand{\eean}{\end{eqnarray*}}

\newcommand{\tcr}[1]{\textcolor{red}{#1}}
\newcommand{\tcb}[1]{\textcolor{blue}{#1}}
\newcommand{\tcg}[1]{\textcolor{green}{#1}}

\newcommand{\tcc}[1]{\textcolor{cyan}{#1}}
\newcommand{\tcm}[1]{\textcolor{magenta}{#1}}
\definecolor{darkorange}{rgb}{.6,.2,.0}
\newcommand{\tcdo}[1]{\textcolor{darkorange}{#1}}
\definecolor{darkgreen}{rgb}{0.0,0.7,0.0}
\newcommand{\tcdg}[1]{\textcolor{darkgreen}{#1}}
\newcommand{\tcp}[5]{\definecolor{#1}{rgb}{#2,#3,#4}\textcolor{#1}{#5}}
\begin{document}


\title{\large\textbf{On the Kinematics of Protein Folding}}


\author{Sean Cahill}
\email{locky@cs.unm.edu}
\affiliation{Department of Computer Science,
University of New Mexico, Albuquerque, NM 87131}
\author{Michael Cahill}
\email{s4mcahill@usuhs.mil}
\affiliation{School of Medicine, 
Uniformed Services University, 
Bethesda, Maryland 20814}
\author{Kevin Cahill}
\email{cahill@unm.edu}
\affiliation{Department of Physics and Astronomy, 
University of New Mexico, Albuquerque, NM 87131}

\date{\today}

\begin{abstract}
\noindent
\textbf{Abstract:} We offer simple solutions to 
three kinematic problems that
occur in the folding of proteins.
We show how to construct suitably local
elementary Monte Carlo moves, 
how to close a loop, and
how to fold a loop without
breaking the bond that closes it.
\end{abstract}


\maketitle

{\flushleft{\textbf{Three Kinematic Problems}\label{3kp}}}\\
\\
Monte Carlo searches for the low-energy states of proteins
pose three related kinematic problems:
\begin{enumerate}
\item How does one design suitably local
elementary Monte Carlo moves?  
\item How does one configure a main-chain loop
between two fixed points?
\item How does one fold a loop without 
breaking any of its bonds?
\end{enumerate}
The choice of elementary 
localized moves may be almost as important 
as the choice of the energy function.
The loop problems occur if one adds or deletes 
residues in a backbone strand
in order to model a homologous or mutant 
protein from a known x-ray structure.
They also occur if one has a loop
that is unresolved by x-ray crystallography
or a primary sequence with two cysteines 
that might form a disulfide bond.
\\
{\flushleft{\textbf{Local Moves}}}\\
\\
The positions \tcr{\(\vec r_i\)} of the atoms
of a protein are local coordinates, but
they are subject to constraints.
The dihedral angles, \tcb{\(\phi_i\)}
and \tcb{\(\psi_i\)},
describe the state of a protein more efficiently,
but they are not local coordinates;
a change in a dihedral angle 
near the center of a protein
rotates half of the molecule, 
moving distant atoms
farther than nearby ones.
Such thrashing violates the
conservation of angular momentum
and of energy, and engenders steric clashes.
Real proteins do not thrash;
they wriggle.
So if one uses the dihedral angles
as coordinates, then one must
craft elementary moves
that are suitably local.
\par
How does one combine rotations
about dihedral bonds 
so that the net motion is suitably local?
This problem was addressed by G\={o}
and Scheraga~\cite{Go1970}
and has since been 
discussed in many papers on 
proteins~\cite{Burkert1982,Bruccoleri1985,Palmer1991,Elofsson1995}
and polymers~\cite{Schatzki1965,Helfand1971,Skolnick&Helfand1980,Helfand&Wasserman&Weber1980,Helfand&Wasserman&Weber1981,Helfand&Wasserman&Weber&Skolnick&Runnels1981,WeberHelfandWasserman1983,Helfand1984,Dodd1993,Leontidis1994,Dinner2000,Kolossvary2001}.
But rotations are complicated.
They are \(3\times3\) orthogonal
matrices with elements
that are sines and cosines 
of the relevant angles.
The nonlinearity of these 
trigonometric functions has held back
progress on this problem.
\par
Yet every smooth function becomes linear
when examined at a small-enough scale.
Rotations of infinitesimally small angles
are linear functions of those angles.
Linear algebra is relatively simple.
\par
The change \tcr{\(\vec dr\)} in the position \tcr{\(\vec r\)}
of an atom due to a rotation 
by a small angle \tcr{\(\epsilon\)} about 
a bond axis represented by the unit vector \tcb{\(\hat b\)} is 
the cross-product of \(\tcm{\epsilon} \, \tcb{\hat b}\)
with the vector to the point \tcr{\(\vec r\)}
from any point \tcb{\(\vec c\)}
on the axis 
\beq
\tcr{\vec dr} = 
\tcm{\epsilon} \, \tcb{\hat b} \times ( \tcr{\vec r} - \tcb{\vec c} ).
\label{dr=}
\eeq
So the change \tcr{\(\vec dr\)} 
due to \(n\) rotations by the small angles 
\(\tcm{\epsilon_i}\)
about the bonds \tcb{\(\hat b_i\)} is the sum
\bea
\tcr{\vec dr} & = & \sum_{i=1}^n 
\tcm{\epsilon_i} \, \tcb{\hat b_i} 
\times ( \tcr{\vec r} - \tcb{\vec c_i} )
\nonumber \\
& = & \mbox{} \sum_{i=1}^n 
\tcm{\epsilon_i} \tcb{\hat b_i} \times 
( \tcr{\vec r} - \tcp{ac}{.2}{.6}{.2}{\vec a} 
+ \tcp{ac}{.2}{.6}{.2}{\vec a} - \tcb{\vec c_i} ) 
\label{dr=sum}\\
& = & \left( \sum_{i=1}^n 
\tcm{\epsilon_i} \, \tcb{\hat b_i} \right) 
\times (\tcr{\vec  r} - \tcp{ac}{.2}{.6}{.2}{\vec a} )
+ \sum_{i=1}^n 
\tcm{\epsilon_i} \, \tcb{\hat b_i} \times 
( \tcp{ac}{.2}{.6}{.2}{\vec a} - \tcb{\vec c_i} ) 
\nonumber
\eea
which is a rotation and a translation.
The point \(\tcp{ac}{.2}{.6}{.2}{\vec a}\) 
is entirely arbitrary; a convenient choice
is the average position 
\(\tcp{ac}{.2}{.6}{.2}{\vec a} = (1/n)\sum_i \tcb{\vec c_i}\)
of the points
\(\tcb{\vec c_i}\).
\par
The rotation is less local than the translation
because its effect is proportional to
the length of the vector 
\((\tcr{\vec  r} - \tcp{ac}{.2}{.6}{.2}{\vec a} )\)\@.
But it is easy to make the net rotation
\beq
\left( \sum_{i=1}^n 
\tcm{\epsilon_i} \, \tcb{\hat b_i} \right) 
\times (\tcr{\vec  r} - \tcp{ac}{.2}{.6}{.2}{\vec a} )
\label{rotation}
\eeq
vanish.  
A set of \(n\) vectors \(\tcb{\hat b_i}\)
is said to be \emph{linearly dependent} if 
there are coefficients \( \tcm{\epsilon_i} \)
such that
\beq
\sum_{i=1}^n \tcm{\epsilon_i} \, \tcb{\hat b_i} = \vec 0 .
\label{angle0}
\eeq
Any set of \(n\) vectors \( \tcb{\hat b_i} \)
in a space of \(d\) dimensions
is linearly dependent if \(n > d \)\@.
Since the bond vectors \(\tcb{\hat b_i}\)
are three dimensional,
any four or more 
are linearly dependent.
So if we use at least 4 bond vectors \(\tcb{\hat b_i}\),
then we always may find angles \(\tcm{\epsilon_i}\)
that make the sum (\ref{angle0}) vanish.
\par
We may find these angles by performing
a singular-value decomposition.
Every \(n \times m\) real matrix \(\tcb{B}\) may be written
as the product of an \( n \times n \) 
orthogonal matrix \(\tcc{U}\),
an \( n \times m \) matrix \(\Sigma\),
and an \(m \times m\) orthogonal matrix \(\tcr{V}\)
in the singular-value decomposition
\beq
\tcb{B} = \tcc{U} \, \Sigma \, \tcr{V}.
\label{svd}
\eeq
The matrix \(\Sigma\) is zero except
for its diagonal matrix elements,
which are the non-negative
singular values of the matrix \(\tcb{B}\)\@.
\par
To find small angles \(\tcm{\epsilon_i}\),  such that 
\beq
\sum_{i=1}^n \tcm{\epsilon_i} \, \tcb{\hat b_i} = \vec 0 ,
\label{angle02}
\eeq
we set \(n=4\) and form a \(3 \times 4\) matrix \(\tcb{B}\)
whose columns are the \(4\) bond vectors \(\tcb{\hat b_i}\)\@.
Its singular-value decomposition is
\bea
\tcb{\lefteqn{\tcb{B} = \pmatrix{b_{11}&b_{12}&b_{13}&b_{14}\cr
                  b_{21}&b_{22}&b_{23}&b_{24}\cr
                  b_{31}&b_{32}&b_{33}&b_{34}}}}\\
& = & \tcc{U} \pmatrix{s_1&0&0&0\cr
                          0&s_2&0&0\cr
                          0&0&s_3&0} 
\pmatrix{\tcr{V_{11}}&\tcr{V_{12}}&\tcr{V_{13}}&\tcr{V_{14}}\cr
            \tcr{V_{21}}&\tcr{V_{22}}&\tcr{V_{23}}&\tcr{V_{24}}\cr
            \tcr{V_{31}}&\tcr{V_{32}}&\tcr{V_{33}}&\tcr{V_{34}}\cr
            \tcr{V_{41}}&\tcr{V_{42}}&
            \tcr{V_{43}}&\tcr{V_{44}}}.\nonumber
\label{Asvd}
\eea
Because the matrices \(\tcc{U}\) and \(\tcr{V}\)
are orthogonal, their rows and columns
are orthonormal vectors.
In particular
\beq
\sum_{k=1}^4 \tcr{V_{ik} \, V_{4k}} = \delta_{i4}.
\label{Vorthonormal}
\eeq
So if we take the small angles to be 
\(\tcm{\epsilon_i} = x \, \tcr{V_{4i}}\), 
where \(x\) is a scale factor,
then 
the orthonormality (\ref{Vorthonormal})
of the rows of \(\tcr{V}\) will imply
\[
\tcr{V} \tcm{\epsilon} = 
\pmatrix{\tcr{V_{11}}&\tcr{V_{12}}&\tcr{V_{13}}&\tcr{V_{14}}\cr
         \tcr{V_{21}}&\tcr{V_{22}}&\tcr{V_{23}}&\tcr{V_{24}}\cr   
         \tcr{V_{31}}&\tcr{V_{32}}&\tcr{V_{33}}&\tcr{V_{34}}\cr
         \tcr{V_{41}}&\tcr{V_{42}}&\tcr{V_{43}}&\tcr{V_{44}}}
\pmatrix{\tcm{x\,V_{41}}\cr\tcm{x\,V_{42}}\cr
            \tcm{x\,V_{43}}\cr\tcm{x\,V_{44}}}
= \pmatrix{0\cr0\cr0\cr x} 
\] 
and so
\beq
\tcb{B} \tcm{\epsilon} = 
\sum_{i=1}^4\tcb{\hat b_i} \, \tcm{\epsilon_i} = 
\tcc{U} \pmatrix{s_1&0&0&0\cr
                 0&s_2&0&0\cr
                 0&0&s_3&0} \pmatrix{0\cr0\cr0\cr x} = \vec 0 .
\label{angle03}
\eeq
\par
\textsc{Lapack}~\cite{LAPACK} is stable tested software 
that solves many problems of linear algebra.
Its subroutine dgesvd performs
singular-value decompositions in
double-precision arithmetic.
The call to
\[
\mathrm{
dgesvd(\textrm{'N'},\textrm{'A'},3,4,\tcb{B},3,S,U,1,\tcr{V},4,WORK,402,INF)}
\]
returns the matrix \(\tcr{V}\)\@.
The small angles \(\tcm{\epsilon_i}\) 
may be taken to be
\beq
\tcm{\epsilon_i} = x \, \tcr{V_{4i}}
\label{epsilons}
\eeq
in which \(x\) is a random number 
in the range \(-\delta < x < \delta \),
and \( \delta \) is small enough
for the small-angle approximations
(\ref{dr=}--\ref{dr=sum}) to be valid.
We used \(\delta = 0.0125\)\@. 
WORK is a double-precision array
of dimension LWORK, here taken to be 402\@.
If the call is successful,
then INF is returned as zero. 
\par
\begin{figure}
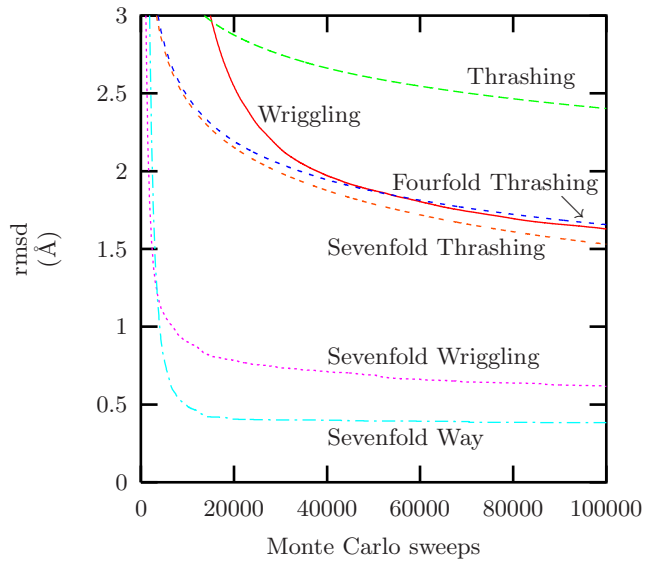

\centering
\input 16PK7f3
\caption{For the protein phosphoglycerate kinase,
\textsc{16pk}, the lines trace the mean values of the
rmsd for 10 runs guided by algorithms
that respectively use \tcdg{thrashing},
\tcb{fourfold thrashing}, \tcr{wriggling},
\tcdo{sevenfold thrashing}, \tcm{sevenfold wriggling},
and the \tcc{sevenfold way}.}
\label{16pk7f3fig}
\end{figure}
\par
We use the word \textit{wriggling\/}
to denote this way of canceling
the non-local effects of rotations.
We tested our wriggling algorithm 
in Monte Carlo simulations of protein folding
against an algorithm in which successive dihedral angles
were varied independently, \textit{thrashing\/},
and also against one in which the
dihedral angles were varied in groups
of four, \textit{fourfold thrashing\/}.
To separate kinematic from dynamic issues,
we used as a nearly perfect but
highly artificial energy function,
the rmsd between the main-chain atoms
of our wriggling protein and 
those of its pdb file.
\par
Because of our use of the rmsd
as an energy function,
the proteins of our simulations are phantoms;
they can pass through themselves.
A real but approximate energy function
would reject all moves into excluded volume;
it therefore would reject many thrashing moves 
because of their large-scale motions.
The use of the rmsd in our tests  
deprives wriggling of one of its key advantages
over thrashing, namely that its localized
motions are less likely to involve steric clashes. 
Thus the utility of wriggling in simulations
with real energy functions may be greater 
than is indicated by our tests.
\par
In our tests of various move sets, 
we let each algorithm fold 10
highly denatured coils of
phosphoglycerate kinase (\textsc{16pk}.pdb, 415 aa),
five of which were stretched. 
The algorithm made a list of the rotatable bonds
which we took to be all the main-chain N-C\(_\alpha\)
and C\(_\alpha\)-C\(^\prime\) bonds,
except for the N-C\(_\alpha\) bonds of the prolines.
In \textsc{16pk}.pdb, there were \(N_B = 810\)
rotatable bonds.
The thrashing code varied all \(N_B\) 
main-chain dihedral angles
in each Monte Carlo sweep,
but the wriggling and fourfold-thrashing codes,
which apply their moves to groups of four bonds at a time,
started with bonds 1--4 and ended 
with bonds \((N_B -3)\)--\(N_B\)\@. 
After 100,000 Monte Carlo sweeps,
the wriggling, thrashing,
and fourfold-thrashing algorithms 
respectively reduced their mean rmsd's to
\(1.63 \pm 0.02\), \(2.40 \pm 0.02\), and \(1.66 \pm 0.03\)~\AA\@.
In Fig.~\ref{16pk7f3fig} the mean rmsd's are plotted
against the number of sweeps. 
These results are similar to 
those we obtained earlier~\cite{Cahill2002}
using C\(_\alpha\) rmsd's.
\par
The use of \(n=4\) bonds \(\tcb{\hat b_i}\)
is the simplest way of canceling 
the highly non-local effects of rotations,
but it is not the best way.
In further tests we found much
lower rmsd's by using \(n=7\) bonds \(\tcb{\hat b_i}\)\@.
We call this \textit{sevenfold wriggling\/}.
The matrix \(\tcb{B}\) is now \(3\times7\), 
each of its 7 columns being a bond vector \(\tcb{\hat b_i}\)\@.
The call is to
\[
\mathrm{
dgesvd(\textrm{'N'},\textrm{'A'},3,7,\tcb{B},3,S,U,1,\tcr{V},7,WORK,460,INF)}.
\]
The angles are given by
\(
\tcm{\epsilon_i} = x \, \tcr{V_{7i}}
\)
where \(|x| < \delta\) is a random number, and
\( \delta \) is small enough
that the small-angle approximations
(\ref{dr=}--\ref{dr=sum}) are valid.
We used \(\delta = 0.0125\)\@.
\par
Sevenfold wriggling dropped the mean rmsd for \textsc{16pk}
to \(0.62 \pm 0.02\) \AA,
as shown in Fig.~\ref{16pk7f3fig}.
We also experimented with using more than seven bonds:
\(n = 8\) gave 0.62 \AA; \(n=9\) gave 0.70 \AA;
\(n=10\) gave 0.76 \AA; and \(n=20\) gave 1.30 \AA\@. 
Sevenfold thrashing gave a mean rmsd of
\(1.53 \pm 0.03\) \AA\@.
\par
By using seven or more bonds,
we may cancel not only the net rotation
\beq
\sum_{i=1}^n 
\tcm{\epsilon_i} \, \tcb{\hat b_i} = 0
\label{rottobecanceled}
\eeq
but also the net translation
\beq
\sum_{i=1}^n 
\tcm{\epsilon_i} \, \tcb{\hat b_i} \times 
( \tcp{ac}{.2}{.6}{.2}{\vec a} - \tcb{\vec c_i} ) = 0 .
\label{transtobecanceled}
\eeq
To do this,
we write these two conditions 
in terms of the six-vectors
\beq
\tcb{s_i} = \pmatrix{\tcb{\hat b_i} \cr 
\tcb{\hat b_i} \times 
( \tcp{ac}{.2}{.6}{.2}{\vec a} - \tcb{\vec c_i} )}
\label{six}
\eeq
as
\beq
\sum_{i=1}^n \tcm{\epsilon_i} \, \tcb{s_i} = 0 .
\label{six=0}
\eeq
Because any 7 or more 6-vectors \(\tcb{s_i}\) are
linearly dependent,
such small angles \(\tcm{\epsilon_i}\) always exist
if at least 7 bonds \(\tcb{\hat b_i}\) are used.
We call such moves \textit{strictly local wriggling\/}.
The matrix \(\tcb{B}\) now is \(6\times7\),
each of its 7 columns being a 6-vector \(\tcb{s_i}\),
and the call is to
\[
\mathrm{
dgesvd(\textrm{'N'},\textrm{'A'},6,7,\tcb{B},6,S,U,1,\tcr{V},7,WORK,850,INF)}.
\]
\par
One might think (as we did) that
strictly local wriggling
is the ideal way to fold a protein.
But although it is well suited to our third kinematic problem,
the folding of a closed loop, 
it is very slow because it leaves 
the first and last backbone atoms unmoved to first
order in the small angles \(\tcm{\epsilon_i}\).
What does work well is the use 
of sevenfold wriggling and strictly local wriggling
on alternate Monte Carlo sweeps along the protein.
We call this technique the \textit{sevenfold way\/}.
It reduced the mean rmsd of \textsc{16pk} 
to \(0.38 \pm 0.02\) \AA,
as shown in Fig.~\ref{16pk7f3fig}.
\begin{figure}
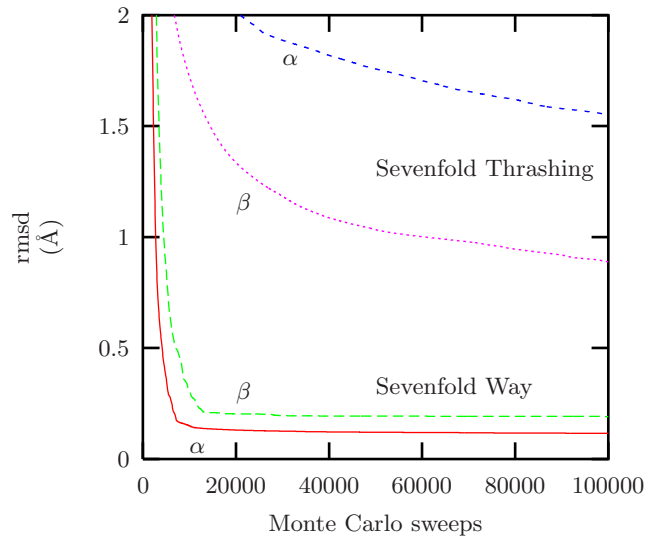

\centering
\input 1A6Mab7ft
\caption{For the protein myoglobin, \textsc{1a6m},
the lines trace the mean values of the
rmsd for 10 runs 
starting from a long \(\alpha\)-helix
and from a long \(\beta\)-strand
guided by \tcb{sevenfold} \tcm{thrashing} 
or the \tcg{sevenfold} \tcr{way}.}
\label{1A6Mab7ftfig}
\end{figure}
\begin{figure}
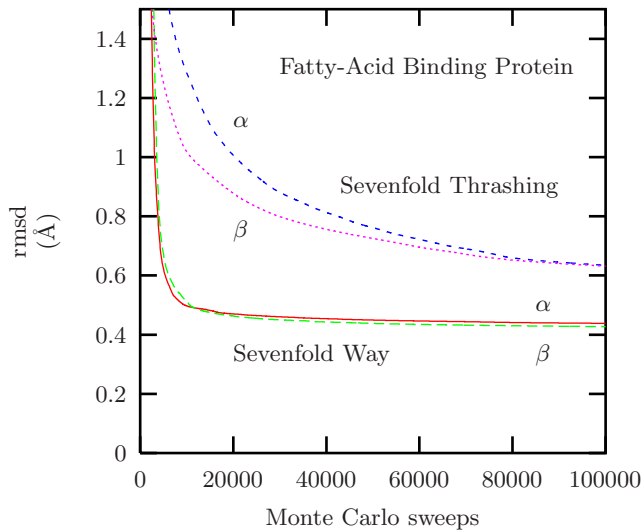

\centering
\input 1HMRab7ft
\caption{For the human muscle 
fatty-acid binding protein, \textsc{1hmr},
the lines trace the mean values of the
rmsd for 10 runs 
starting from a long \(\alpha\)-helix
and from a long \(\beta\)-strand
guided by \tcb{sevenfold} \tcm{thrashing} 
or the \tcr{sevenfold} \tcg{way}.}
\label{1HMRab7ftfig}
\end{figure}
\par
We also compared the sevenfold way
to sevenfold thrashing on two
other globular proteins --- sperm-whale myoglobin
(\textsc{1a6m}.pdb, 151 aa)
and human muscle fatty-acid binding protein
(\textsc{1hmr}.pdb, 131 aa)\@.
On these shorter proteins,
we used the sevenfold way on 
all sets of seven contiguous bonds,
fourfold wriggling on the three final
sets of four contiguous bonds,
and allowed arbitrary rotations
about the last three bonds.
For each protein, we made two denatured starting
configurations, one that was a single long straight
\(\beta\)-strand and one that was a single long straight
\(\alpha\)-helix.  
In 10 runs of 100,000 sweeps,
sevenfold thrashing
reduced the mean main-chain rmsd's of the long \(\alpha\)-helix
and of the long \(\beta\)-strand of \textsc{1a6m}
to \(1.55 \pm 0.01\)
and \(0.89 \pm 0.03 \)~\AA\ respectively.
The sevenfold-way did much better,
respectively reducing the rmsd's  
to \(0.12 \pm 0.02\) 
and \(0.19 \pm 0.05\)~\AA,
as shown in Fig.~\ref{1A6Mab7ftfig}\@. 
In the case of \textsc{1hmr},
the mean rmsd's of the long \(\alpha\)-helix
and of the long \(\beta\)-strand 
respectively were reduced 
in 10 runs of 100,000 sweeps
to \(0.63 \pm 0.02\) 
and \(0.63 \pm 0.01\)~\AA\ 
by sevenfold thrashing
and to  \(0.44 \pm 0.02\)
and \(0.43 \pm 0.03\)~\AA\ 
by the sevenfold way,
as shown in Fig.~\ref{1HMRab7ftfig}\@.
\par
\begin{figure}
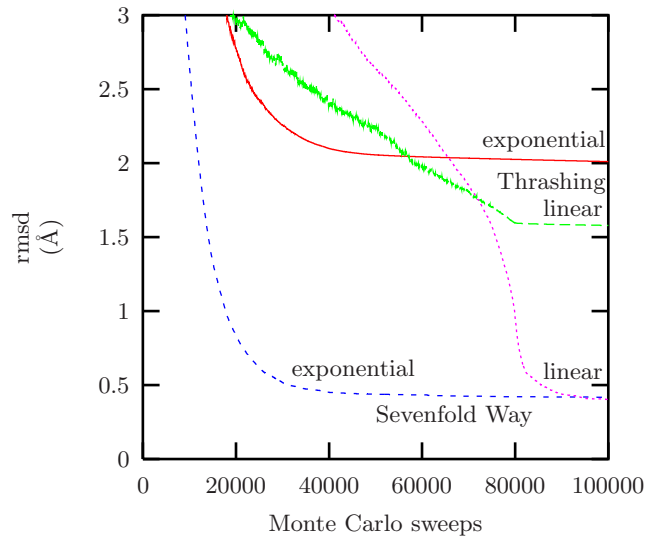

\centering
\input 16PKT
\caption{For the protein \textsc{16pk},
the lines trace the mean values of the
rmsd for 10 runs guided by algorithms
that respectively use thrashing
and the sevenfold way with
linear and exponential cooling.}
\label{16pkTfig}
\end{figure}

Because our artificial energy function,
the rmsd, is not directly related
to an energy, 
our fixed-temperature simulations
were carried out 
at zero temperature.
We did however test the sevenfold way
against thrashing 
in runs with simulated annealing.
In these Monte Carlo simulations,
the temperature dropped 
either linearly or exponentially
with the sweep number from 
a very high \(T\) at the start
of the simulation to \(T=0\)
at 80,000 sweeps.  These runs
finished with 20,000 sweeps at \(T=0\)\@.
As shown in Fig.~\ref{16pkTfig}, the sevenfold way
reduced the mean rmsd of \textsc{16pk} to \(0.42 \pm 0.01\) \AA\ 
with exponential cooling
and to \(0.40 \pm 0.02\) \AA\ 
with linear cooling.
Thrashing reduced 
the mean rmsd to \(2.01 \pm 0.03\) \AA\ 
with exponential cooling
and to \(1.58 \pm 0.15\) \AA\ 
with linear cooling.
We did not try entropic sampling~\cite{Lee1993}\@.
\par
In these algorithms,
the changes in the dihedral angles
are small so that
the small-angle approximations
(\ref{dr=}--\ref{dr=sum}) are valid;
atoms are displaced 
in each move by no more
than about 0.05 \AA\@.
Such small moves might seem 
to imply a slow rate of
convergence to minimum-energy states.
But globular proteins typically
are 50 \AA\ or less in diameter, 
and so in 1,000 moves an atom 
could traverse the protein.
To test whether larger moves
would give faster convergence,
we made 10 thrashing runs with
a cutoff 40 times bigger than 
our usual \(\delta = 0.0125\)\@.
These runs with large-angle thrashing reduced 
the mean main-chain rmsd 
in 100,000 sweeps 
to \(3.69 \pm 0.06\) \AA,
which is to be compared to \(2.40 \pm 0.02\) \AA\
with \(\delta = 0.0125\)\@.
So bigger moves may not mean faster convergence,
perhaps because in a partially folded protein,
smaller moves are more likely than larger ones
to lower the energy.
\\
{\flushleft{\textbf{To Close a Loop}\label{close}}}\\
\\
When modeling a homologous or mutant protein 
or an unresolved loop,
one must configure a main chain
between two fixed points.
This problem also arises if one has a main chain 
with two cysteine residues, and one needs
to make a disulfide bridge between them,
forming a loop.
\par
Let us consider the case
of a loop with a disulfide bridge.
Provided the strand of backbone is long
enough, we may change the dihedral angles
of the residues of the strand between the cysteines
so as to move the \(\beta\)-carbon and the \(\gamma\)-sulfur
of the second cysteine into the required positions
opposite those of the first cysteine,
which is held fixed.
Let \(\tcr{\vec C_{\beta 0}}\) and \(\tcr{\vec S_{\gamma 0}}\) 
be the points to which 
the \(\beta\)-carbon and the \(\gamma\)-sulfur
should be moved, and let
\(\tcr{\vec C_{\beta}}\) and \(\tcr{\vec S_{\gamma}}\) 
be their present locations.
\par
We have seen in Eq.(\ref{dr=sum}) that
several small rotations amount to
a net rotation and a net translation.
We may choose the small angles of the 
rotations so as to correctly orient the 
\(\tcr{\vec C_{\beta} - \vec S_{\gamma}}\) 
bond and to move it to the right position.
\par
The required translation is
\beq
\tcr{\vec t} = \frac{1}{2} \left(
\tcr{\vec S_{\gamma 0} + \vec C_{\beta 0} 
- \vec S_\gamma - \vec C_\beta} \right) .
\label{t}
\eeq 
The axis of the required rotation is
\beq
\tcr{ \vec x } = 
\left( \tcr{\vec S_{\gamma} - \vec C_{\beta}} \right)
\times \left( \tcr{\vec S_{\gamma 0} - \vec C_{\beta 0}} \right),
\label{x}
\eeq
and the tangent of its angle \(\theta\) is
the ratio of the length \(\|\tcr{ \vec x }\|\)
of this cross-product to the corresponding
scalar-product
\beq
\tcr{ d } = 
\left( \tcr{\vec S_{\gamma} - \vec C_{\beta}} \right)
\cdot \left( \tcr{\vec S_{\gamma 0} - \vec C_{\beta 0}} \right),
\label{d}
\eeq
that is,
\(\tan \theta = \|\tcr{ \vec x }\|/d \)\@.
So the required angle of rotation is
\beq
\tcr{ \vec \theta } = 
\mathrm{atan2}\,(\|\tcr{\vec x}\|,\tcr{d}) \; \tcr{\hat x}.
\label{theta}
\eeq
Thus we must perform \(n \ge 6\) small rotations,
each of angle \(\tcm{\epsilon_i}\)
about bond \(\tcb{\hat b_i}\), 
so that
the net rotation is 
\beq
\tcr{ \vec \theta } = 
\sum_{i=1}^n 
\tcm{\epsilon_i} \, \tcb{\hat b_i} 
\label{rottobecanceled2}
\eeq
and the net translation is 
\beq
\tcr{\vec t} =
\sum_{i=1}^n 
\tcm{\epsilon_i} \, \tcb{\hat b_i} \times 
( \tcp{ac}{.2}{.6}{.2}{\vec a} - \tcb{\vec c_i} ) .
\label{transtobecanceled2}
\eeq
These two conditions may be written
in terms of the 6-vectors
\beq
\tcb{s_i} = \pmatrix{\tcb{\hat b_i} \cr 
\tcb{\hat b_i} \times 
( \tcp{ac}{.2}{.6}{.2}{\vec a} - \tcb{\vec c_i} )}
\label{six2}
\eeq
as
\beq
\sum_{i=1}^n \tcm{\epsilon_i} \, \tcb{s_i} = 
\pmatrix{\tcr{\vec \theta}\cr
         \tcr{\vec t}}      .
\label{six=t}
\eeq
We form a matrix \(\tcb{B}\) 
whose first \(n\) columns are the
6-vectors \(\tcb{s_i}\)
and whose last column is the 6-vector
\( \pmatrix{\tcr{\mbox{} - \vec \theta},
&\tcr{\!\!\!\!\mbox{} - \vec t}} \)\@.
A call to dgesvd returns the matrix \(\tcr{V}\),
and the small angles \(\tcm{\epsilon_i}\)
are given by a suitably safe version 
of \(\tcm{\epsilon_i} = \tcr{V(j,i)/V(j,n+1)}\),
where \(j > 6\) is the row index 
with the largest value of \(|\tcr{V(j,n+1)}|\)\@.
In our applications of this transport algorithm,
we set \(n\) equal to the number of 
rotatable main-chain bonds of the loop.
\par
\begin{figure}
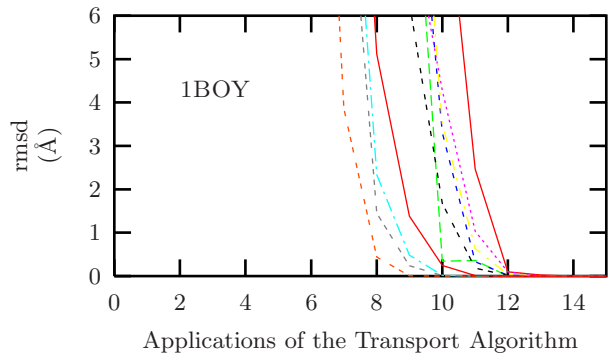

\centering
\input 1BOYcl
\caption{The transport algorithm 
closed the 44-bond loop of the protein \textsc{1boy}
in 14 sweeps.}
\label{1BOYclfig}
\end{figure}
\begin{figure}
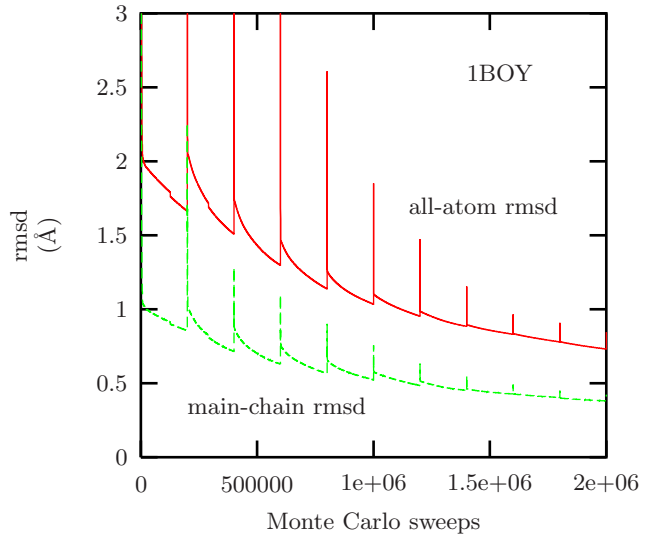

\centering
\input 1BOYfl
\caption{The sevenfold way needed about
a million sweeps to reduce the \tcg{main-chain} 
and \tcr{all-atom} rmsd's 
of the loop of 44 rotatable bonds in \textsc{1boy} 
below 1 \AA\@.
The spikes are caused by the application
of the transport algorithm,
which closes the loop after
every 200,000 sweeps.}
\label{1BOYflfig}
\end{figure}

The extracellular domain of human tissue factor 
(\textsc{1boy}.pdb, 219 aa) has two disulfide bonds.
The one between residues 186 and 209 closes a loop
that has 44 rotatable main-chain bonds.
The transport algorithm with \(n = 44\) 
closed this loop to less than \(0.0001\)\AA\ 
in 14 sweeps,
as shown in Fig.~\ref{1BOYclfig}.
\\
{\flushleft{\textbf{To Fold a Loop}}}\\
\\
When the transport algorithm closes
a loop, the loop may well be of quite
high energy with steric conflicts.
It is therefore necessary to vary
the conformation of the loop without breaking it.  
Strictly local wriggling moves
are well suited to this task.
But we have found that the sevenfold way
combined with the transport algorithm
does a better job.
We used the sevenfold way 
on all sets of seven contiguous bonds
and applied the transport algorithm
after every 200,000 sweeps of the sevenfold way. 
As an energy function, we used
the all-atom rmsd
between the atoms of the folding
loop and those of the pdb file.
In 10 runs of 2,000,000 sweeps
from fully extended coils
of the 44-bond loop of \textsc{1boy},
this process reduced the loop's mean
all-atom rmsd to 
\(0.85 \pm 0.14\) \AA\ and
its mean main-chain
rmsd to \(0.42 \pm 0.06\) \AA,
as shown in Fig.~\ref{1BOYflfig}\@.
\begin{figure}
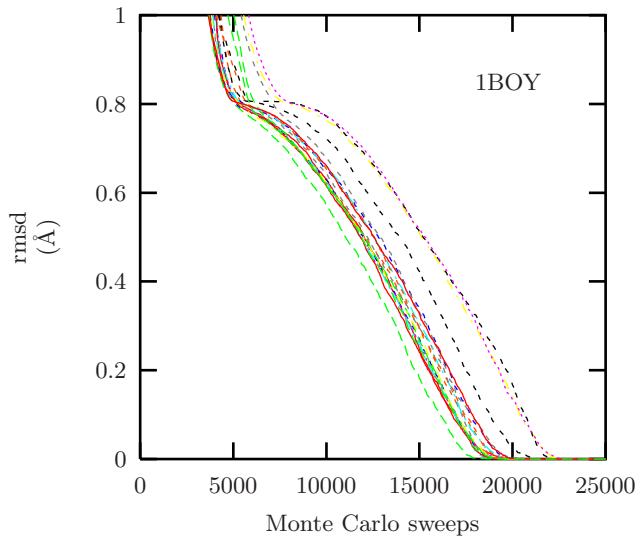

\centering
\input 1BOYNfl025
\caption{The n-fold way reduced to zero the all-atom rmsd 
of the loop of 44 rotatable bonds in \textsc{1boy} 
in all 20 runs of 25,000 Monte Carlo sweeps.}
\label{1BOYNfl025fig}
\end{figure}
\begin{figure}
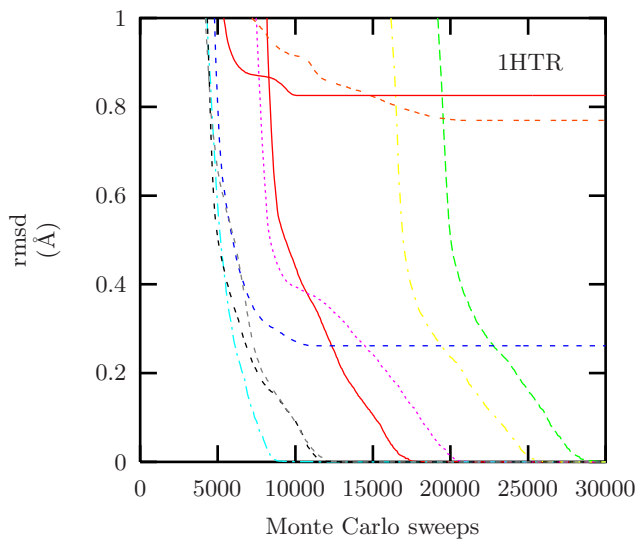

\centering
\input 1HTR44Nmfl025
\caption{The n-fold way reduced to zero the all-atom rmsd 
of the loop of 60 rotatable bonds in human progastricsin, 
\textsc{1htr}.pdb,
in 7 of 10 runs of 30,000 Monte Carlo sweeps.}
\label{1HTR44Nmfl025fig}
\end{figure}
\par
The sevenfold way needed a million sweeps
to reduce the mean all-atom rmsd 
of the 44-bond loop in \textsc{1boy}
to less than 1 \AA\@. 
One may do better by letting the program 
randomly choose how many bonds, \(7 \le n \le 44\),
to fold and which row, \(7 \le j \le n\), 
of the matrix \(V\) of the singular-value
decomposition (\ref{svd})
to use for the angles 
\(\tcm{\epsilon_i} = x \, \tcr{V_{ji}}\).
A further advantage is gained by allowing
suitable moves of the last six bonds,
as was done for the proteins \textsc{1a6m}
and \textsc{1hmr}\@.
In 10 runs of 20,000 sweeps,
this algorithm reduced the mean all-atom rmsd
of the 44-bond loop of \textsc{1boy}
to \(0.66 \pm 0.22\) \AA\@.
The large standard error arose
from the fact the rmsd got stuck
at 1.326 \AA\ in 5 of the runs,
but went to less than 0.02 \AA\ in the other 5.
\par
Hoping to send the rmsd of every run to zero,
we introduced a new Monte Carlo move.
When a protein has folded to
a globular form close to the
native structure,
a net translation may disrupt
the folding protein as much as or more
than a net rotation.  So we added
a new move in which the net translation,
but not the net rotation, is set to zero:
\beq
\tcr{\vec t} = 
\sum_{i=1}^n \tcm{\epsilon_i} \, \tcb{\hat b_i} \times 
( \tcp{ac}{.2}{.6}{.2}{\vec a} - \tcb{\vec c_i} )  
= 0      .
\label{three=t=0}
\eeq
One may enforce this condition by
forming a matrix \(\tcb{B}\) whose \(n \ge 4\)
columns are the vectors \( \tcb{\hat b_i} \times 
( \tcp{ac}{.2}{.6}{.2}{\vec a} - \tcb{\vec c_i} ) \)\@.
The resulting \textit{n-fold way} 
on successive sweeps sets to zero
net translations, net rotations,
and both the net translations and the net rotations.
It reduced the rmsd
of the 44-bond loop of \textsc{1boy} to zero in
20 out of 20 runs in fewer than 25,000 sweeps,
as shown in Fig.~\ref{1BOYNfl025fig}. 
\par
The aspartyl protease
human progastricsin
(\textsc{1htr}.pdb, 329 aa) has three 
disulfide bonds.
The one between residues 251 and 284 
closes a loop that has 60 rotatable bonds. 
We denatured and stretched that
loop and then used 
the n-fold way to fold it.
By limiting the number \(n\) 
of simultaneously rotated bonds
to 44, which by default was true
in the case of \textsc{1boy}, we found
that we could drive the rmsd
to zero in 7 out of 10 runs,
as shown in Fig.~\ref{1HTR44Nmfl025fig}\@.
In simulations with \(n\) set equal
to 20, 30, 40, and 50,
we found that the rmsd went to zero
in 4 or 5 out of 10 runs.
\\
{\flushleft{\textbf{Summary}}}\\
\\
We have presented simple solutions to 
three kinematic problems that
occur in the folding of proteins.
We have shown how to construct suitably local
elementary Monte Carlo moves, 
how to close a loop, and
how to fold a loop without
breaking any of its bonds.
\par
In future work,
we intend to determine whether
these kinematic algorithms 
improve the efficiency 
of finite-temperature Monte Carlo searches 
guided by realistic energy functions
with implicit solvation and excluded volume.
\\
{\flushleft{\textbf{Acknowledgments}}}\\
\\
We have benefited from talking 
with S.~Atlas, D.~Baker, B.~Brooks, G.~Cahill,
P.~Cahill, E.~Coutsias, L.~deEcheandia, K.~Dill, 
D.~Dix, S.~Fellini, H.~Frauenfelder,
K.~Frost, N.~van~Gulick, U.~Hansmann,
G.~Herling, D.~Lanar, A.~Parsegian, W.~Saslow,
C.~Schwieters, C.~Seok, P.~J.~Steinbach, C.~Strauss
and R.~Venable.
Some of this work was done
while one of us (KC) was
on sabbatical in 
the Center for Molecular Modeling
of the Center for Information Technology
of the National Institutes of Health.
We should like to thank
Ken Dill and Peter J.~Steinbach,
respectively,
for the hospitality they each 
extended to S.~C.\ at UCSF and K.~C.\ at NIH\@.
The UNM RAC 
provided a grant for software.
This study used the high-performance computational capabilities 
of the Biowulf/LoBoS3 cluster at the National Institutes of Health, 
Bethesda, Md. 
Some of our computations were
done by the computers of  
the Albuquerque High-Performance
Computing Center.
\bibliography{mc,coh,cs,polymers,proteins}

\begin{thebibliography}{20}
\expandafter\ifx\csname natexlab\endcsname\relax\def\natexlab#1{#1}\fi
\expandafter\ifx\csname bibnamefont\endcsname\relax
  \def\bibnamefont#1{#1}\fi
\expandafter\ifx\csname bibfnamefont\endcsname\relax
  \def\bibfnamefont#1{#1}\fi
\expandafter\ifx\csname citenamefont\endcsname\relax
  \def\citenamefont#1{#1}\fi
\expandafter\ifx\csname url\endcsname\relax
  \def\url#1{\texttt{#1}}\fi
\expandafter\ifx\csname urlprefix\endcsname\relax\def\urlprefix{URL }\fi
\providecommand{\bibinfo}[2]{#2}
\providecommand{\eprint}[2][]{\url{#2}}

\bibitem[{\citenamefont{G\={o} and Scheraga}(1970)}]{Go1970}
\bibinfo{author}{\bibfnamefont{N.}~\bibnamefont{G\={o}}} \bibnamefont{and}
  \bibinfo{author}{\bibfnamefont{H.~A.} \bibnamefont{Scheraga}},
  \bibinfo{journal}{Macromolecules} \textbf{\bibinfo{volume}{3}},
  \bibinfo{pages}{178} (\bibinfo{year}{1970}).

\bibitem[{\citenamefont{Burkert and Allinger}(1982)}]{Burkert1982}
\bibinfo{author}{\bibfnamefont{U.}~\bibnamefont{Burkert}} \bibnamefont{and}
  \bibinfo{author}{\bibfnamefont{N.~L.} \bibnamefont{Allinger}},
  \emph{\bibinfo{title}{Molecular Mechanics}}, ACS Monographs
  (\bibinfo{publisher}{American Chemical Society},
  \bibinfo{address}{Washington, D.C.}, \bibinfo{year}{1982}).

\bibitem[{\citenamefont{Bruccoleri and Karplus}(1985)}]{Bruccoleri1985}
\bibinfo{author}{\bibfnamefont{R.}~\bibnamefont{Bruccoleri}} \bibnamefont{and}
  \bibinfo{author}{\bibfnamefont{M.}~\bibnamefont{Karplus}},
  \bibinfo{journal}{Macromolecules} \textbf{\bibinfo{volume}{18}},
  \bibinfo{pages}{2767} (\bibinfo{year}{1985}).

\bibitem[{\citenamefont{Palmer and Scheraga}(1991)}]{Palmer1991}
\bibinfo{author}{\bibfnamefont{K.~A.} \bibnamefont{Palmer}} \bibnamefont{and}
  \bibinfo{author}{\bibfnamefont{H.~A.} \bibnamefont{Scheraga}},
  \bibinfo{journal}{J.\ Computational Chemistry} \textbf{\bibinfo{volume}{12}},
  \bibinfo{pages}{505} (\bibinfo{year}{1991}).

\bibitem[{\citenamefont{Elofsson et~al.}(1995)\citenamefont{Elofsson, Le~Grand,
  and Eisenberg}}]{Elofsson1995}
\bibinfo{author}{\bibfnamefont{A.}~\bibnamefont{Elofsson}},
  \bibinfo{author}{\bibfnamefont{S.~M.} \bibnamefont{Le~Grand}},
  \bibnamefont{and}
  \bibinfo{author}{\bibfnamefont{D.}~\bibnamefont{Eisenberg}},
  \bibinfo{journal}{Proteins: Structure, Function, and Genetics}
  \textbf{\bibinfo{volume}{23}}, \bibinfo{pages}{73} (\bibinfo{year}{1995}).

\bibitem[{\citenamefont{Schatzki}(1965)}]{Schatzki1965}
\bibinfo{author}{\bibfnamefont{T.~F.} \bibnamefont{Schatzki}},
  \bibinfo{journal}{Polymer Preprints (American Chemical Society, Division of
  Polymer Chemistry)} \textbf{\bibinfo{volume}{6}}, \bibinfo{pages}{646}
  (\bibinfo{year}{1965}).

\bibitem[{\citenamefont{Helfand}(1971)}]{Helfand1971}
\bibinfo{author}{\bibfnamefont{E.}~\bibnamefont{Helfand}},
  \bibinfo{journal}{Journal of Chemical Physics} \textbf{\bibinfo{volume}{54}},
  \bibinfo{pages}{4651} (\bibinfo{year}{1971}).

\bibitem[{\citenamefont{Skolnick and Helfand}(1980)}]{Skolnick&Helfand1980}
\bibinfo{author}{\bibfnamefont{J.}~\bibnamefont{Skolnick}} \bibnamefont{and}
  \bibinfo{author}{\bibfnamefont{E.}~\bibnamefont{Helfand}},
  \bibinfo{journal}{Journal of Chemical Physics} \textbf{\bibinfo{volume}{72}},
  \bibinfo{pages}{5489} (\bibinfo{year}{1980}).

\bibitem[{\citenamefont{Helfand et~al.}(1980)\citenamefont{Helfand, Wasserman,
  and Weber}}]{Helfand&Wasserman&Weber1980}
\bibinfo{author}{\bibfnamefont{E.}~\bibnamefont{Helfand}},
  \bibinfo{author}{\bibfnamefont{Z.~R.} \bibnamefont{Wasserman}},
  \bibnamefont{and} \bibinfo{author}{\bibfnamefont{T.~A.} \bibnamefont{Weber}},
  \bibinfo{journal}{Macromolecules} \textbf{\bibinfo{volume}{13}},
  \bibinfo{pages}{526} (\bibinfo{year}{1980}).

\bibitem[{\citenamefont{Helfand
  et~al.}(1981{\natexlab{a}})\citenamefont{Helfand, Wasserman, and
  Weber}}]{Helfand&Wasserman&Weber1981}
\bibinfo{author}{\bibfnamefont{E.}~\bibnamefont{Helfand}},
  \bibinfo{author}{\bibfnamefont{Z.~R.} \bibnamefont{Wasserman}},
  \bibnamefont{and} \bibinfo{author}{\bibfnamefont{T.~A.} \bibnamefont{Weber}},
  \bibinfo{journal}{Polymer Preprints (American Chemical Society, Division of
  Polymer Chemistry)} \textbf{\bibinfo{volume}{22}}, \bibinfo{pages}{279}
  (\bibinfo{year}{1981}{\natexlab{a}}).

\bibitem[{\citenamefont{Helfand
  et~al.}(1981{\natexlab{b}})\citenamefont{Helfand, Wasserman, Weber, Skolnick,
  and Runnels}}]{Helfand&Wasserman&Weber&Skolnick&Runnels1981}
\bibinfo{author}{\bibfnamefont{E.}~\bibnamefont{Helfand}},
  \bibinfo{author}{\bibfnamefont{Z.~R.} \bibnamefont{Wasserman}},
  \bibinfo{author}{\bibfnamefont{T.~A.} \bibnamefont{Weber}},
  \bibinfo{author}{\bibfnamefont{J.}~\bibnamefont{Skolnick}}, \bibnamefont{and}
  \bibinfo{author}{\bibfnamefont{J.~H.} \bibnamefont{Runnels}},
  \bibinfo{journal}{Journal of Chemical Physics} \textbf{\bibinfo{volume}{75}},
  \bibinfo{pages}{4441} (\bibinfo{year}{1981}{\natexlab{b}}).

\bibitem[{\citenamefont{Weber et~al.}(1983)\citenamefont{Weber, Helfand, and
  Wasserman}}]{WeberHelfandWasserman1983}
\bibinfo{author}{\bibfnamefont{T.~A.} \bibnamefont{Weber}},
  \bibinfo{author}{\bibfnamefont{E.}~\bibnamefont{Helfand}}, \bibnamefont{and}
  \bibinfo{author}{\bibfnamefont{Z.~R.} \bibnamefont{Wasserman}},
  \emph{\bibinfo{title}{Simulation of Polyethylene}}
  (\bibinfo{publisher}{American Chemical Society},
  \bibinfo{address}{Washington, DC}, \bibinfo{year}{1983}),
  \bibinfo{type}{molecular-based study of fluids}~\bibinfo{chapter}{20}, pp.
  \bibinfo{pages}{487--500}, no. \bibinfo{number}{204} in \bibinfo{series}{ACS
  Advances in Chemistry Series}.

\bibitem[{\citenamefont{Helfand}(1984)}]{Helfand1984}
\bibinfo{author}{\bibfnamefont{E.}~\bibnamefont{Helfand}},
  \bibinfo{journal}{Science} \textbf{\bibinfo{volume}{226}},
  \bibinfo{pages}{647} (\bibinfo{year}{1984}).

\bibitem[{\citenamefont{Dodd et~al.}(1993)\citenamefont{Dodd, Boone, and
  Theodorou}}]{Dodd1993}
\bibinfo{author}{\bibfnamefont{L.~R.} \bibnamefont{Dodd}},
  \bibinfo{author}{\bibfnamefont{T.~D.} \bibnamefont{Boone}}, \bibnamefont{and}
  \bibinfo{author}{\bibfnamefont{D.~N.} \bibnamefont{Theodorou}},
  \bibinfo{journal}{Molecular Physics} \textbf{\bibinfo{volume}{78}},
  \bibinfo{pages}{961} (\bibinfo{year}{1993}).

\bibitem[{\citenamefont{Leontidis et~al.}(1994)\citenamefont{Leontidis,
  de~Pablo, Laso, and Suter}}]{Leontidis1994}
\bibinfo{author}{\bibfnamefont{E.}~\bibnamefont{Leontidis}},
  \bibinfo{author}{\bibfnamefont{J.~J.} \bibnamefont{de~Pablo}},
  \bibinfo{author}{\bibfnamefont{M.}~\bibnamefont{Laso}}, \bibnamefont{and}
  \bibinfo{author}{\bibfnamefont{U.~W.} \bibnamefont{Suter}},
  \bibinfo{journal}{Advances in Polymer Science}
  \textbf{\bibinfo{volume}{116}}, \bibinfo{pages}{283} (\bibinfo{year}{1994}).

\bibitem[{\citenamefont{Dinner}(2000)}]{Dinner2000}
\bibinfo{author}{\bibfnamefont{A.~R.} \bibnamefont{Dinner}},
  \bibinfo{journal}{Journal of Computational Chemistry}
  \textbf{\bibinfo{volume}{21}}, \bibinfo{pages}{1132} (\bibinfo{year}{2000}).

\bibitem[{\citenamefont{Kolossv\`{a}ry and Keser\^{u}}(2001)}]{Kolossvary2001}
\bibinfo{author}{\bibfnamefont{I.}~\bibnamefont{Kolossv\`{a}ry}}
  \bibnamefont{and} \bibinfo{author}{\bibfnamefont{G.~M.}
  \bibnamefont{Keser\^{u}}}, \bibinfo{journal}{Journal of Computational
  Chemistry} \textbf{\bibinfo{volume}{22}}, \bibinfo{pages}{21}
  (\bibinfo{year}{2001}).

\bibitem[{\citenamefont{Anderson et~al.}(1999)\citenamefont{Anderson, Bai,
  Bischof, Blackford, Demmel, Dongarra, Du~Croz, Greenbaum, Hammarling,
  McKenney et~al.}}]{LAPACK}
\bibinfo{author}{\bibfnamefont{E.}~\bibnamefont{Anderson}},
  \bibinfo{author}{\bibfnamefont{Z.}~\bibnamefont{Bai}},
  \bibinfo{author}{\bibfnamefont{C.}~\bibnamefont{Bischof}},
  \bibinfo{author}{\bibfnamefont{S.}~\bibnamefont{Blackford}},
  \bibinfo{author}{\bibfnamefont{J.}~\bibnamefont{Demmel}},
  \bibinfo{author}{\bibfnamefont{J.}~\bibnamefont{Dongarra}},
  \bibinfo{author}{\bibfnamefont{J.}~\bibnamefont{Du~Croz}},
  \bibinfo{author}{\bibfnamefont{A.}~\bibnamefont{Greenbaum}},
  \bibinfo{author}{\bibfnamefont{S.}~\bibnamefont{Hammarling}},
  \bibinfo{author}{\bibfnamefont{A.}~\bibnamefont{McKenney}},
  \bibnamefont{et~al.}, \emph{\bibinfo{title}{LAPACK Users' Guide}}
  (\bibinfo{publisher}{SIAM}, \bibinfo{address}{Philadelphia, PA},
  \bibinfo{year}{1999}), \bibinfo{edition}{3rd} ed., \bibinfo{note}{available
  on-line at http://www.netlib.org/lapack/lug/lapack\_lug.html}.

\bibitem[{\citenamefont{Cahill et~al.}(2002)\citenamefont{Cahill, Cahill, and
  Cahill}}]{Cahill2002}
\bibinfo{author}{\bibfnamefont{M.}~\bibnamefont{Cahill}},
  \bibinfo{author}{\bibfnamefont{S.}~\bibnamefont{Cahill}}, \bibnamefont{and}
  \bibinfo{author}{\bibfnamefont{K.}~\bibnamefont{Cahill}},
  \bibinfo{journal}{The Biophysical Journal} \textbf{\bibinfo{volume}{82}},
  \bibinfo{pages}{2665} (\bibinfo{year}{2002}), \eprint{cond-mat/0108218}.

\bibitem[{\citenamefont{Lee}(1993)}]{Lee1993}
\bibinfo{author}{\bibfnamefont{J.}~\bibnamefont{Lee}},
  \bibinfo{journal}{Physical Review Letters} \textbf{\bibinfo{volume}{71}},
  \bibinfo{pages}{211} (\bibinfo{year}{1993}).

\end{thebibliography}

\end{document}